# SCMA based resource management of D2D communications for maximum sum-revenue

Linglin Kong[1], Li Ling[1], and Xu Zhang[2]
[1]*School of Information Science and Technology, Fudan University,* Shanghai, China
[2]*Cainiao Network Technology Co., Ltd, Alibaba Group,* Hangzhou, China
E-mail:[1]{17210720133, lingli}@fudan.edu.cn, [2] zhangxu_hit@163.com

*Abstract*—The device-to-device (D2D) communication is one of the promising technologies of the future Internet of Things (IoT), but its security-related issues remain challenging. The block-chain is considered to be a secure and reliable distributed ledger, so we can treat the device user equipment (D-UE) request for the reusing resources of cellular user equipment (C-UE) as a transaction and put it into a transaction pool, then package the record into the block-chain. In this paper, we study the D2D communication resource allocation scheme based on sparse code multiple access (SCMA). Firstly, the system's interference model and block-chain-based transaction flow are analyzed. Then we propose the optimization problem so that C-UE can get the maximum revenue by sharing its resources to D-UE. This problem is NP-hard, so we propose a heuristic algorithm based on semi-definite relaxation (SDR) programming to solve it. Finally, the performance of the proposed algorithm is verified by simulation of different system parameters.

## I. INTRODUCTION

With the increase of smart devices and the development of network technology, the IoT has become a hot research topic. There are more and more communication devices with higher requirements in the network, as well as the efficiency and security of communication. 5G communication can make communications have lower latency and more device accessing [1]. Through D2D communication, two devices can communicate directly without passing through the base station (BS), which can reduce the number of hops and time delay, which is more suitable for the localization scenario of the IoT [2]. Mobile edge computing (MEC) moves the computing tasks of the network from the center to the edge users so that D2D communications can alleviate the pressure on BS and improve system efficiency. In [3][4], the non-orthogonal multiple access (NOMA) can allow the network access more users by introducing non-orthogonal resource blocks (RB) more than orthogonal multiple access (OMA), although the D-UE reusing the resources of C-UE would introduce interference, but we can utilize the control of the quality of service (QoS) of each device to work properly [5].

Most of the previous researches focused on resource scheduling of centralized BS to maximize the communication throughput of C-UE and D-UE in the scenario [6]. However, they do not take into account the fairness of communication, that is, the C-UE who are shared resources do not receive any compensation, so this paper considers the optimization problem of C-UE revenue in D2D communications. Recently, the block-chain is considered to be a secure and reliable distributed accounting technology [7]. It has been widely used in finance and computer science, but there is still no in-depth research in mobile communication. The block-chain maintained by all nodes may partly replace the function of the centralized database, and also conforms to the partial decentralization feature of D2D communication and MEC. In this paper, a D2D pair initiates the reusing of C-UE can be seen as a transaction which would be recorded in the block-chain, thus ensuring the security of each transaction. In addition, in order to ensure the fairness of communications for all users, the user initiates a transaction can pay a certain tip (like the 'Gas' in Ethereum [8]) to obtain a higher transaction priority, which can make resource allocation more efficient.

The rest of this paper is organized as follows: system model is proposed in Section II, including cellular hybrid networks, multiple access, and block-chain-based transaction flow. The optimization problem is described in Section III. Section IV presents the corresponding heuristic algorithm based on SDR. Section V illustrates the simulation results. Section VI concludes this paper.

## II. SYSTEM MODEL

### A. Cell Model

As shown in Fig. 1, this paper considers a single cell scenario, including C-UE and D-UE, each of which is considered as a node in a block-chain. To simplify the model, we assume that all D2D communications are occurred in the same cell to avoid interference from another BS. Where $\boldsymbol{C} = \{1,2,...,N\}$ is a set of C-UE, and $\boldsymbol{D} = \{1,2,...,K\}$ is a set of D-UE.

In order to enable more devices in the system, we utilize the SCMA as the C-UE's non-orthogonal multiple access method to the BS [9]. The SCMA maps $\log_2 M$ bits into an $M$-dimensions codebook by an encoder, where $M$ is the codeword length. Therefore, the SCMA coding mapping of user $k$ can be expressed as $\vec{s_k} = S_k(\mathbf{b})$, where $\mathbf{b}$ is the input binary bit stream and $S_k(*)$ is the mapping function of user $k$. $\vec{s_k}$ is a $M$-dimensions sparse codeword vector, and the number of non-zero elements is $M_C < M$, so the maximum number of codebooks is $J = C_{M_C}^M$. If $R$ is the number of resource blocks, we can define the overload factor $OF = J/R$ for this mapping. Obviously, in SCMA, $OF > 1$ and $OF = 1$ in OMA. Fig. 2 shows an example of SCMA encoder.

### B. Transactions over Block-chains

Based on the characteristics of D2D communication, this paper proposes a block-chain-based D2D communication transaction scheme. Taking the communication between a D2D pair as an example, the specific process of the transaction flow is shown in Fig. 3.

Here are some details: ① D-UE transmitter (Tx) sends a request of reuse to the C-UE; ② C-UE forwards the request to the D-UE receiver (Rx); ③ D-UE Rx agrees to this request; ④ C-UE transfers the acknowledgement of D-UE Rx to the D-UE Tx; ⑤ C-UE releases the resource to both D-UE Tx and D-UE Rx, and starts timing the reuse; ⑥ Either of pairs notifies the C-UE that the communication is ended, and the timer stops work; ⑦ Clearing the transaction, then put it into the trading pool. If the balance of the paying party is enough, put the transaction into the full payment trading pool, otherwise, the partial payment trading pool, change into another after they pay total fees; ⑧ The transactions in full payment trading pool are packaged into block-chains periodically, ultimately implementing security features.

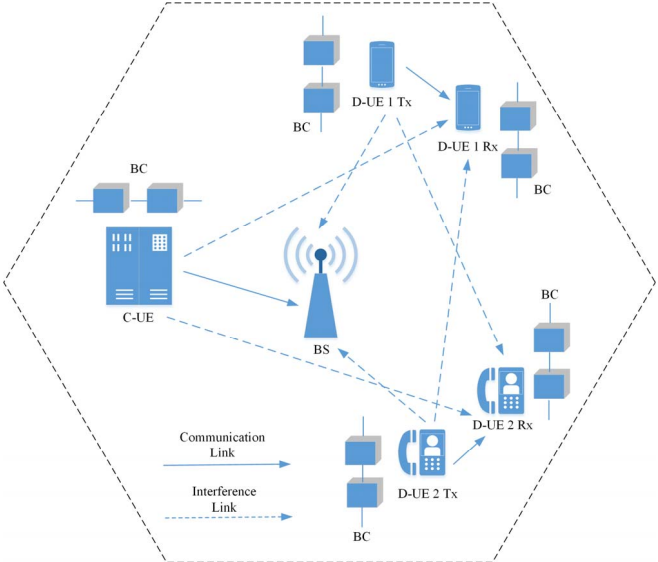

Fig. 1. Channel sharing and interference model of mixed cell with BC

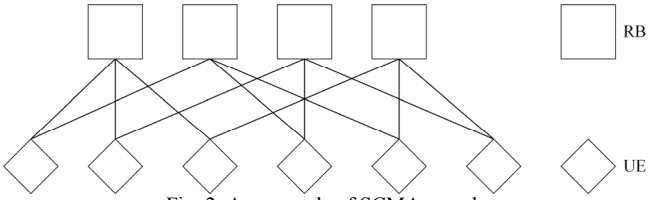

Fig. 2. An example of SCMA encoder

In this paper, we assume the D2D pair pay an optional tip to the C-UE in addition to the cost of leasing resources when initiating the transaction, which is proportional to their final transaction amount, so that they can make the transaction a higher priority. We assume that there are two types of D2D pairs: central users and edge users, which are related to their distance $Distance_{ik}$ to C-UE $k$, and edge users can pay more tip to obtain C-UE resources. Because the physical distance between a D2D pair is generally close, they can be approximated as a D-UE, so that $\boldsymbol{D'} = \{1,2,…,K'\}$ is a set of equivalent D2D pairs, and $K' = C_K^2$. Therefore, the $m$-th time a D2D pair (equivalent D-UE $i$) pays to C-UE $k$ because of reusing is:

$$R(D_i, C_k) = T_m(D_{ik})(1 + \beta_{ik}) \qquad (1)$$

where $T_m(D_{ik})$ is the time at which the D-UE $i$ reuses the C-UE $k$'s resource, and $\beta_{ik}$ is the optional tip rate.

## III. PROBLEM FORMULATION

### A. Optimization Problem

Since there is resource reusing between the D-UE and C-UE, which would cause interference to the C-UE. By introducing a block-chain-based compensation mechanism, C-UE is more willing to lease their own resources, but need to guarantee their own QoS.

It is assumed that the BS can get the channel state information (CSI) of each device in the cell, and the UEs can periodically send a request to the BS to query the information, the signal overhead is enough small. Considering that the BS has a stronger anti-interference capability, the D-UE is applied to reuse the uplink channels of C-UE and BS. The goal of this problem is to maximize the revenues of C-UE. Based on the analysis of communication interference within the cell and (1), the problem can be raised as follows:

$$P0: \max \sum_{i=1}^{K'} \sum_{k=1}^{N} \sum_m T_m(D_{ik})(1+\beta_{ik})$$

$$s.t.\ C1: \gamma_{CUE-k} = \frac{P_k^C h_{ck,bs}}{\sum_{i=1}^{K'} P_k^D h_{di,bs} x_{ik} + \sum_{l=1,l\neq k}^{N} P_l^C h_{ck,cl} y_{kl} + N_0} \geq SINR_{\min}^C, \forall k \in \boldsymbol{C}$$

$$C2: \gamma_{DUE-i} = \frac{P_i^D h_{di,dj}}{\sum_{k=1}^{N} P_k^C h_{ck,bs} x_{ik} + \sum_{j=1,j\neq i}^{K'} P_j^D h_{dj,di} + N_0} \geq SINR_{\min}^D, \forall i \in \boldsymbol{D'}$$

$$C3: 0 \leq P_k^C \leq P_{\max}^C, \forall k \in \boldsymbol{C}$$

$$C4: 0 \leq P_i^D \leq P_{\max}^D, \forall i \in \boldsymbol{D'}$$

$$C5: 0 \leq N \leq N_{RB} \cdot OF$$

where $\gamma_{CUE-k}, \gamma_{DUE-i}$ are the actual SINR of C-UE $k$ and D-UE $i$, respectively, $h_{a,b}$ denotes the channel gain between $a$ and $b$, fading environment is ignored here. The $P_k^C, P_i^D$ are the actual power of C-UE $k$ and D-UE $i$, respectively. Furthermore, $x_{ik}$ or $y_{kl}$ is a binary system reusing coefficient: if D-UE $i$ reuses the RB of C-UE $k$, then $x_{ik}=1$, otherwise 0. if C-UE $k$ and C-UE $l$ reuse the same RB, $y_{kl}=1$, otherwise 0. The $P_{max}^C, P_{max}^C, SINR_{min}^C$ and $SINR_{min}^D$ are the maximum operating power and minimum SINR requirement for C-UE and D-UE, respectively. In addition, $N_0$ is the Gaussian white noise, and $N_{RB}$ is the number of RBs in the cell.

Among them, C1 and C2 ensure that C-UE and D-UE meet their QoS requirements respectively. C3 and C4 indicate that the working powers of C-UE and D-UE do not exceed their maximum working powers; C5 indicates that the maximum number of C-UE that can be accessed in a cell does not exceed the product of the number of RBs in the cell and overload factor.

### B. Problem decomposition

Due to the suddenness and uncertainty of communication, it is difficult to solve the above problem because it is tough to make a valid estimate of $\sum_m T_m(D_i)$ in a period of time. To simplify the problem, we assume that the maximum time $T_{max} = \tau$ for each D2D communication, which also meets the block-chain generation cycle, that is, the block-chain confirms transactions occurring in the previous cycle. Therefore, within time $T$, the D-UE can have up to $S_{max} = T/\tau$ time slots (TS) for communication. A binary variable $\delta(D_i, \tau_s)$ is defined: if the D-UE $i$ has a communication in the $s$-th TS, $\delta(D_i, \tau_s) =1$, otherwise $\delta(D_i, \tau_s) =0$. Where $\delta(D_i, \tau_s) \sim N(\mu, \sigma^2)$, that is, the probability of communication occurring in the TS obeys a normal Gaussian distribution, and each time the number of active D-UE is $K''$. In addition, the C3~C5 are related to the physical characteristics of the UEs and BS, so they do not affect the final result during the optimization, which can be omitted.

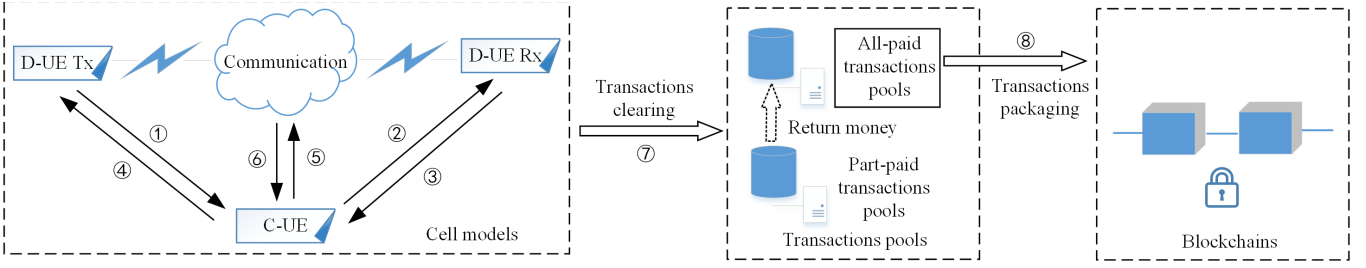

Fig.3. Transactions over block-chains

Considering that in each TS, the location of the device is fixed, so $h_{a,b}$ is a constant without considering the effect of fading environment. Let the initial SINR of C-UE $k$ and the D-UE $i$ in the $s$-th TS be $\gamma_0(SINR_{k,s}^C)$ and $\gamma_0(SINR_{i,s}^D)$, respectively. And the increment caused by the access of the D-UE $i$ for SINR of C-UE $k$ and D-UE $j$ is $-\gamma_i(SINR_{k,s}^C)$ and $-\gamma_j(SINR_{k,s}^D)$, respectively. So $P0$ can be translated into:

$$P1: \max \sum_{i=1}^{K^*} \sum_{k=1}^{N} \sum_{m} \tau\delta(D_{ik},\tau_s)(1+\beta_{ik})$$

$$s.t.\ C1': \gamma_{CUE-k} = \gamma_0(SINR_{k,s}^C) -$$
$$\sum_{i=1}^{K^*} \gamma_i(SINR_{k,s}^C)\delta(D_{ik},\tau_s) \geq SINR_{\min}^C,\ \forall k \in C$$
$$C2': \gamma_{DUE-i} = \gamma_0(SINR_{i,s}^D) -$$
$$\sum_{j=1,j\neq i}^{K^*} \gamma_j(SINR_{k,s}^D)\delta(D_{ik},\tau_s) \geq SINR_{\min}^D,\ \forall i \in D'$$

## IV. SOLUTIONS TO PROBLEM

Firstly, considering the case of single TS $s$ and single C-UE $k$, the $P1$ can be transformed into:

$$P2: \max \sum_{i=1}^{K^*} \tau\delta(D_{ik},\tau_s)(1+\beta_i)$$
$$s.t.\ C1',\ C2'$$

where $P2$ with $C1'$ can be equivalent to a knapsack problem, denoted as $P3$, which is np-hard, and $C2'$ as a constraint.

First, define $P3$ with follows by introducing some new parameters:

$$\begin{cases} \sum_{i=1}^{K''} \tau\delta(D_{ik},\tau_s)(1+\beta_i) = \Delta_i(\Phi)\delta(D_{ik},\tau_s) \\ y_i = 2\delta(D_{ik},\tau_s) - 1 \\ \gamma_i(SINR_{k,s}^C) = h_i \\ \gamma_0(SINR_{k,s}^C) - SINR_{\min}^C = C \end{cases} \quad (2)$$

$$P3: \max \frac{1}{2}\sum_{i=1}^{K^*} \Delta_i(\Phi) y_i + \frac{1}{2}\sum_{i=1}^{K^*} \Delta_i(\Phi)$$
$$s.t.\ C6': \frac{1}{2}\sum_{i=1}^{K^*} h_i y_i \leq C - \frac{1}{2}h_i$$
$$C7': y_i \in \{-1,1\}$$

If $y$ is an optimal solution for $P3$, $H = \sum_{i=1}^{K''} h_i$, $h_{max} = max\{h_i\}$, assume $2C \leq H$, there will be:

$$\sum_{i=1}^{K^*} h_i y_i \geq 2(C-h_{\max}) - H \quad (3)$$

$$\left(\sum_{i=1}^{K^*} h_i y_i\right)^2 \leq \left(2(C-h_{\max}) - H\right)^2 \rightarrow$$
$$\sum_{i\neq j=1}^{K^*} h_i h_j y_i y_j \leq \left(2(C-h_{\max}) - H\right)^2 - \sum_{i=1}^{K^*} h_i^2 \quad (4)$$

setting these as new constraints to $P3$:

$$P3': \max \frac{1}{2}\sum_{i=1}^{K^*} \Delta_i(\Phi) y_i + \frac{1}{2}\sum_{i=1}^{K^*} \Delta_i(\Phi)$$
$$s.t.\ C6',\ C7',\ (3),\ (4)$$

adding a human variable $y_0$ to further deal with $P3'$:

$$P3'': \max \frac{1}{2}\sum_{i=1}^{K^*} \Delta_i(\Phi) y_i y_0 + \frac{1}{2}\sum_{i=1}^{K^*} \Delta_i(\Phi)$$
$$s.t.\ C6'': \frac{1}{2}\sum_{i=1}^{K^*} h_i y_i y_0 \leq C - \frac{1}{2}h_i$$
$$C7'': y_i \in \{-1,1\}$$
$$C8'': \sum_{i=1}^{K^*} h_i y_i y_0 \geq 2(C-h_{\max}) - H$$
$$C9'': \sum_{i\neq j=1}^{K^*} h_i h_j y_i y_j \leq \left(2(C-h_{\max}) - H\right)^2 - \sum_{i=1}^{K^*} h_i^2$$

Given an optimal solution $y = \{y_1, y_2, \cdots y_n\}$ of $P3'$, the D-UE $i$ who reuses C-UE is denoted as $I(y) = \{i|y_i = 1\}$. Similarly, $y = \{y_0, y_1, \cdots y_n\}$ of $P3''$, the D-UE $i$ who reused C-UE is denoted as $I(y) = \{i|y_i y_0 = 1\}$. If: $\begin{cases} PI = \{I(y) | y \text{ is the optimal solution of } P3'\} \\ PIA = \{IA(y) | y \text{ is the optimal solution of } P3''\} \end{cases}$, $PI = PIA$. In [10], $diag(Y) = e, rank(Y) = 1, Y \geq 0$, $e$ is a n+1 dimensions all-1 column vector. So the SDR for $P3$ is:

$$P3\text{-SDR}: \max \frac{1}{2}\sum_{i=1}^{K^*} \Delta_i(\Phi) Y_{i0} + \frac{1}{2}\sum_{i=1}^{K^*} \Delta_i(\Phi)$$
$$s.t.\ C1\text{-SDR}: \frac{1}{2}\sum_{i=1}^{K^*} h_i Y_{i0} \leq C - \frac{1}{2}h_i$$
$$C2\text{-SDR}: \sum_{i=1}^{K^*} h_i Y_{i0} \geq 2(C-h_{\max})h$$
$$C3\text{-SDR}: \sum_{i,j\neq 1}^{K^*} h_i h_j Y_{i0} \leq \left(2(C-h_{\max}) - H\right)^2 - \sum_{i=1}^{K^*} h_i^2$$
$$C4\text{-SDR}: diag(Y) = e, Y \geq 0$$

TABLE 1. SIMULATION PARAMETERS

| Parameters | Values |
|---|---|
| Radius of cell | 250m |
| Carrier frequency | 2GHz |
| System bandwidth | 20MHz |
| C-UE transmit power | 20dBm |
| D-UE transmit power | 17dBm |
| Noise power spectral density | -174dBm/Hz |

Therefore, the algorithm for solving *P*3 is Algorithm 1. Because this is an np-hard problem that cannot get the optimal solution in polynomial time, this algorithm obtains a suboptimal solution.

**Algorithm 1.** Solutions for *P*3

1: Initialization: $\varepsilon = 0$, the solution $Y$ to *P*3-SDR, $\mu \sim N(0, \rho Y + (1-\rho)I)$, where $0 \leq \rho \leq 1$, $y_i = \begin{cases} 1 & \mu_i \geq 0 \\ -1 & others \end{cases}$, $S^* = \Phi$, $\forall \Phi \subseteq D'$
2: **if** $|\Phi| \leq 3$ and $\sum_{i=1}^{K''} h_i \leq C$, **then**
3:   **if** $C2'$ is feasible, **then**
4:     put D-UE *i* in $\Phi$ into **cluster**1, $f(\Phi_1) = \max f(\Phi)$
5:   **else**
6:     $\varepsilon = 1$
7:   **end if**
8: **else**
9:   **if** $\sum_{i=1}^{K''-3} h_i > C$, **then**
10:     **repeat**: re-send the $S^*$ subscript number, let:
$$\frac{\Delta_{i_1}(\Phi)}{h_{i_1}} \geq \frac{\Delta_{i_2}(\Phi)}{h_{i_2}} \geq \cdots \geq \frac{\Delta_{i_{|S^*|}}(\Phi)}{h_{i_{|S^*|}}}$$
11:     remove $i_{|S^*|}$ from $S^*$, $S^* = \{i_1, i_2, \cdots i_{|S^*|-1}\}$, until $\sum_{i=1}^{K''-3} h_i \leq C$
12:   **else**
13:     $\exists j \in \Phi \setminus S^*$, **repeat**: let $h_i \leq C - \sum_{i=1}^{K''-3} h_i$
14:     **if** $C2'$ is feasible, **then**
15:       put C-UE *j* into **cluster**2 until, $\forall j \in \Phi \setminus S^*$, $h_j > C - \sum_{i=1}^{K''-3} h_i$
16:     **else**
17:       $\varepsilon = 1$
18:     **end if**
19:     $f(\Phi_2) = \max f(\Phi)$
20:   **end if**
21: **end if**
22: output: $f(\Phi) = \max(f(\Phi_1), f(\Phi_2))$

Extending all C-UE, this paper proposes a heuristic algorithm to achieve resource allocation is Algorithm 2:

**Algorithm 2.** Heuristic algorithm for *P*1

1: Initialization: the $\gamma_0(SINR_{k,s}^C)$ of each C-UE *k*
2: **for** $s = 1, \cdots, S_{max}$, **do**
3:   sort C-UE *k* in descending order of $\gamma_0(SINR_{k,s}^C)$
4:   **for** $k = 1, \cdots, N$, **do**
5:     **for** $i = 1, \cdots, K''$, **do**
6:       run Algorithm1
7:       **if** $\varepsilon = 1$, **then**
8:         $k = k + 1$
9:       **end if**
10:       $D' = D' - \Phi$
11:     **end for**
12:   **end for**
13: **end for**

## V. SIMULATION RESULTS

In this simulation, a scenario of a single BS cell is considered, and the system is a hybrid communication network including C-UE and D-UE, both UEs are randomly distributed in the cell. The system channel model considers the provisions of the 3GPP TS 36.104 and 3GPP TR 2_5.943 protocols.

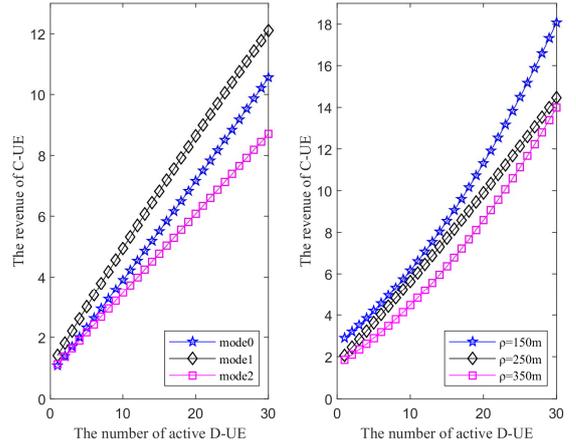

(a)          (b)
Fig. 4. Effect of revenue

We set the SCMA parameter is: the number of codebooks that can be mapped every 4 subcarriers is *J*=6 (see Fig. 2). System modeling parameters are as table 1 shows before.

### A. Effect of revenue

This article applies different modes for verification: *mode0*: all UE apply the same tip rate, which is actually a traditional solution; *mode1*: edge UE need to pay more tip, while central users do not pay the extra tip, $\beta_{ik-center} = 1$, we consider D-UE *i* whose distance to C-UE *k* exceeds $\rho$ as an edge UE, so $\beta_{ik-edge} = Distance_{ik}/\rho > 1$; *mode2*: the tip rate of every D-UE is inversely proportional to the distance to C-UE *k*.

From Fig. 4 (a), we can see that as the number of D-UE increases, the revenue of C-UE also increases, since more active D-UE pay for C-UE. The center D-UE pay more tip in *mode1*, so C-UE gets more revenue than *mode2*. The *mode0* is in between them because there is no difference between all UE in this mode. Further considering the impact of $\beta_{ik}$ on C-UE's revenue, we chose different $\rho$, and its simulation results under *mode1* is shown in Fig. 4 (b). It can be found that as the $\rho$ increases, the number of D-UE that need to pay an extra tip gradually decreases, so the revenue of the C-UE is also less and less.

### B. Effect of capacity

Firstly, we compare the impact of different access methods on C-UE's throughput in Fig.5 (a). Because the SCMA can allow more C-UE to access the system than the OFDMA when the number of RB is the same. When *OF*=1.5, it can be seen that the C-UE reach the maximum throughput when its amount is 45 (the product of the number of RBs and *OF*), and then the throughput cannot continue to increase due to the limited number of RB. In addition, the throughput under SCMA is slightly less than 1.5 times that of OFDMA because a small amount of interference is actually introduced in this access mode. Therefore, the throughput under OFDMA will be slightly higher than SCMA before reaching their maximum throughputs. As can be seen from Fig. 5 (b), the throughput of the D-UE increases as the number of active D-UE increases, and the throughput of the C-UE gradually decrease because more interference is introduced. In addition, the total throughput of the system shows a trend of increasing first and then decreasing, but in general, the introduction of D2D communication can make the system throughput higher.

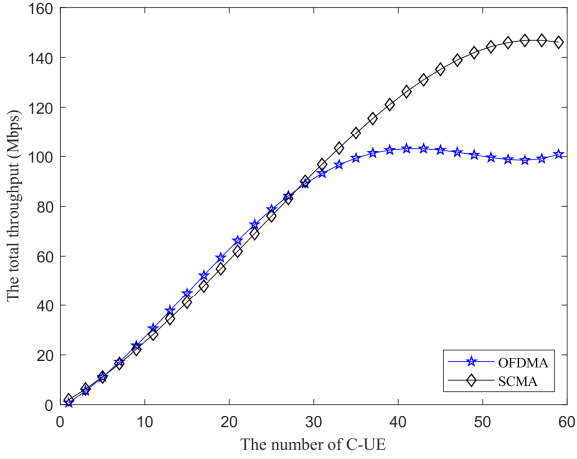

(a)

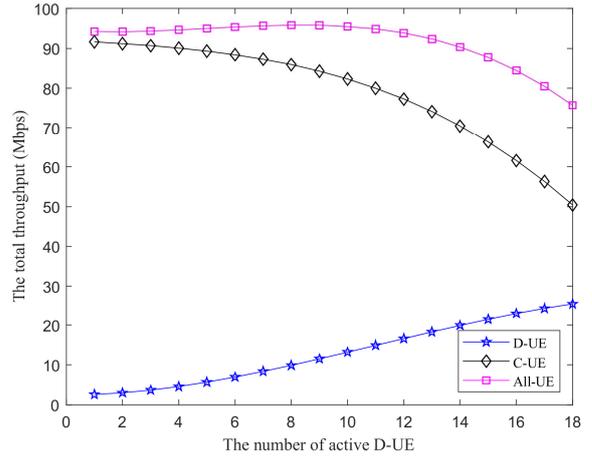

(b)

Fig.5. Effect of capacity

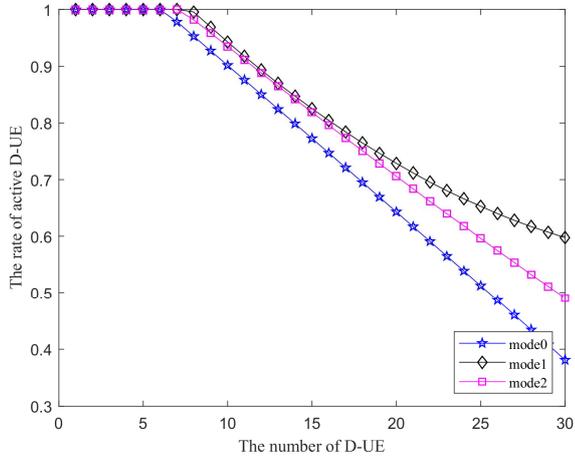

Fig.6. Effect of fairness

*C. Effect of fairness*

Suppose there are 30 C-UE in the system, and the ratio of D-UE active in the communication under *mode0~mode2* and all D-UE is calculated as $\partial = Num_{active}/Num_{all}$. The simulation result is shown in Fig 6. It can be seen that there are more active D-UE access the system in *mode1*, because in *mode2*, when D-UE increase, D-UE who are closer to C-UE are less likely to reuse RB because they pay less. The *mode0*, on the other hand, does not consider the fairness of D-UE with poor channel conditions, so the number of active D-UE is the least.

## VI. CONCLUSION

In this paper, we study the resource allocation problem of D2D communications based on SCMA. In order to maximize the total revenue of C-UE, we propose an optimization problem and a suboptimal solution of its heuristic algorithm based on semi-definite relaxation programming. Simulation results show that our work can indeed provide better performances than traditional solutions. In future work, we will continue to build the block-chain involved in this article. And the feasibility of this algorithm combined with other NOMA schemes would be discussed.